# Spectral characterization of weak coherent state sources based on two-photon interference

Thiago Ferreira da Silva,[1,2,*] Gustavo C. do Amaral,[1] Douglas Vitoreti,[3]
Guilherme P. Temporão,[1] and Jean Pierre von der Weid[1]

[1]*Center for Telecommunications Studies, Pontifical Catholic University of Rio de Janeiro,
Rua Marquês de São Vicente 225 Gávea, Rio de Janeiro, RJ, Brazil*
[2]*Optical Metrology Division, National Institute of Metrology, Quality and Technology,
Av. Nossa Sra. das Graças 50 Xerém, Duque de Caxias, RJ, Brazil*
[3]*Physics Academic Unity, Federal University of Campina Grande,
Rua Aprígio Veloso 882, Campina Grande, PB, Brazil*
*\*Corresponding author: thiago@opto.cetuc.puc-rio.br*

We demonstrate a method for characterizing the coherence function of coherent states based on two-photon interference. Two states from frequency mismatched faint laser sources are fed into a *Hong-Ou-Mandel* interferometer and the interference pattern is fitted with the presented theoretical model for the quantum beat. The fitting parameters are compared to the classical optical beat when bright versions of the sources are used. The results show the equivalence between both techniques.

## 1. Introduction

Weak coherent states (WCSs) are a practical and inexpensive way to probabilistically create single-photon pulses. They are created with a faint laser and are largely employed in quantum cryptography systems for quantum key distribution (QKD) [1]. Due to the probabilistic nature of the number of photons in a time interval for a WCS, there is no way to create a single-photon pulse with certainty so the probabilities of emission of both multi-photon and vacuum pulses must be managed since they are highly correlated. Multi-photon pulses must be avoided in QKD systems, due to the possibility of eavesdropping through a photon-number splitting attack. This is usually accomplished by highly attenuating the source so that the average number of photons per pulse, $\mu$, falls well below 1 [1]. This weak regime bounds the multi-photon to single-photon emission ratio to $\mu/2$, at the cost of highly increasing the vacuum emission probability to $1-\mu$. By making use of the decoy states technique [2-4], however, the value of $\mu$ can be increased to $O(1)$ without compromising the security of the QKD system.

Two-photon interference between single photons was first observed using photon-pairs emitted through spontaneous parametric down conversion (SPDC). By feeding a beamsplitter with identical single photons in its input ports a decrease in the coincidence counts at the outputs occurs due to the photon bunching effect, known as the *Hong-Ou-Mandel* dip [5]. The effect has also been observed with independent SPDC-based sources [6].

When the photons have different frequencies, a quantum beat pattern is expected [7], but the effect cannot be observed unless the coherence time of the single photons is long enough with respect to the detectors timing resolution [8].

Two-photon interference can be observed even when coherent states are employed in a setup where coincidence detections are used to post-select two-photon states from mixed states. Interference between a coherent state and a single-photon has been demonstrated to exhibit non-classical visibility [9]. When two coherent states are used, however, the interference visibility is bounded to 50% for two spatial modes, due to multi-photon emission [9-11]. A superposition of multiple indistinguishable two-photon paths can, however, lead to enhanced visibility values [12].

In this paper we demonstrate a method for the spectral characterization of coherent states in the weak regime based on two-photon interference in a beamsplitter. Two WCS sources, reference and test, are fed into a beamsplitter and the interference pattern is obtained by measuring coincidence counts in a *Hong-Ou-Mandel* (HOM) interferometer. A theoretical model was derived and fits the interference pattern revealing the frequency mismatch and coherence length of the source under characterization. The model considers WCSs expanded up to two photons, so the error is bounded to 1% for an average number of photons per time interval smaller than 0.22. The parameters of the model fit to the interference pattern are compared to the spectrum obtained from the optical beat of bright versions of the optical sources in a photodiode, observed in an electrical spectrum analyzer (ESA). The results show the equivalence between both techniques for different frequency mismatch between optical sources.

## 2. Spectral characterization of WCS sources

The mutual coherence of two WCSs can be obtained with an HOM interferometer, as shown in Fig 1.

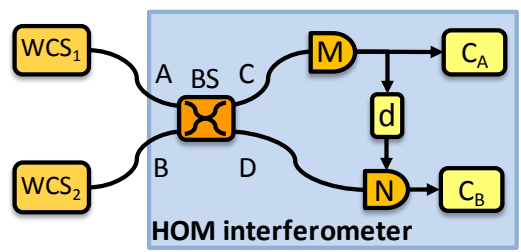

Fig. 1. Method for spectral characterization of WCS$_2$ by two-photon interference with WCS$_1$ using an HOM interferometer based on coincident detections behind a beamsplitter. BS: beamsplitter; M,N: SPDs; d: delay generator; C: pulse counter.

Consider two continuous-wave (CW) WCS sources with identical optical power and parallel states of polarization (SOPs) feeding the input spatial modes, A and B, of a beamsplitter (BS). Two single-photon detectors (SPDs), M and N, are placed each at an output spatial mode of the BS, C and D. The detectors operate in gated mode, with SPD M running with internal gate. Each time SPD M clicks, a voltage pulse is sent to trigger SPD N. Pulse counters (C) are used to acquire the coincident counts between the detectors. The interference pattern is characterized by the coherence time of the sources, σ, and their frequency difference, Δ. This is measured by varying the temporal delay (d) between SPD M and SPD N.
An analytical model, presented in the next section, fits the interference pattern so parameters $\sigma$ and $\Delta$ can be extracted.

## 3. Theoretical model

### A. Coherent states

The coherent state is defined as a superposition of *n*-photon Fock states:

$$|\alpha\rangle = \exp\left(-\frac{|\alpha|^2}{2}\right) \sum_n \frac{\alpha^n}{\sqrt{n!}} |n\rangle \quad (1)$$

where $|\alpha|^2 = \mu$ is the average number of photons in a time interval. The probability of finding *n* photons in a given time interval follows the Poisson distribution, i.e., $P(n|\mu) = \langle n|\alpha^*\alpha|n\rangle = \mu^n exp(-\mu)/n!$. In the weak regime, of small values of $\mu$, the single-photon probability approaches $\mu$. It is interesting to note, however, that coherent states do not reach Fock sates in the asymptotic limit of $\mu$ as it goes to zero.

If two WCS sources with similar values of $\mu$ feed a BS, the probability of finding a pair of Fock states $|m,n\rangle_{A,B}$ at the input modes is given by

$$P(m, n|\mu) = \mu^{m+n} exp(-2\mu)/(m!\,n!) \quad (2)$$

Summing (2) for all combinations of *m* and *n* we note that, for values of μ smaller than 0.22 photons per time interval, the coincidence counts can be described considering only 2 photons (*m* + *n* = 2) at the input of the HOM interferometer with an error smaller than 1%. We will then keep this restriction on our theoretical model as well as the corresponding limitation in the average number of photons per detection gate in our experiments. Since the SPDs are not photon-number resolving, the non-vacuum Fock states are not discriminated in the mixed states. Nevertheless, as long as we only consider the weak regime, the states with more than 2 photons can be disregarded without jeopardizing the validity of the model.

The probability of both sources emitting a single photon simultaneously, $P(1,1|\mu)$ is equal to the probability of any source emitting two photons while the other emits vacuum, $P(2,0|\mu) + P(0,2|\mu)$. This observation will prove to be useful when analyzing the limited visibility of 0.5 for interference between WCSs.

### B. Spatio-temporal modes of wave-packets

Consider a symmetrical optical BS with spatial modes labeled as shown in Fig. 1. We can attribute electric field operators to its input

$$\begin{cases} E_A^+(t) = \xi_A(t)a_A \\ E_B^+(t) = \xi_B(t)a_B \end{cases} \therefore \begin{cases} E_A^-(t) = \xi_A^*(t)a_A^\dagger \\ E_B^-(t) = \xi_B^*(t)a_B^\dagger \end{cases} \quad (3)$$

and the field operators can be described by spatio-temporal modes $\xi_k(t) = \varepsilon(t)exp(-j\varphi_k(t))$, composed by an amplitude $\varepsilon_k(t)$ and a phase $\varphi_k(t)$. Here, the spatial position of the BS has been taken as reference.

The output of the beamsplitter relates to the input fields according to [13,14]

$$\begin{cases} E_C^+(t) = [-jE_A^+(t) + E_B^+(t)]/\sqrt{2} \\ E_D^+(t) = [E_A^+(t) - jE_B^+(t)]/\sqrt{2} \end{cases} \quad (4)$$

### C. Coincidences at the output of the BS

We now analyze the probability of the coincident detection of photons at times $\tau_0$ and $\tau_0 + \tau$ – given by the detection gate window of the SPDs – respectively on the two output modes of the BS. Restricting the analysis up to 2 photons, the possible two-photon input states are $|1,1\rangle_{A,B}$, $|2,0\rangle_{A,B}$, and $|0,2\rangle_{A,B}$. Only parallel-polarized photons are considered here.

In the first case, a single photon comes from each WCS source, and the input state is $|\psi_{in}\rangle = a_A^\dagger a_B^\dagger |0,0\rangle_{A,B}$. By applying the field operators, the coincident detection probability is computed through

$$P_{C,D}^{1,1}(\tau_0,\tau) = \langle\psi_{in}|E_C^-(\tau_0)E_D^-(\tau_0+\tau)E_D^+(\tau_0+\tau)E_C^+(\tau_0)|\psi_{in}\rangle \quad (5)$$

which, through the relationship defined in (3) and (4), leads to [14]

$$P_{C,D}^{1,1}(\tau_0,\tau) = \frac{1}{4}\left|\xi_A(\tau_0+\tau)\xi_B(\tau_0) - \xi_A(\tau_0)\xi_B(\tau_0+\tau)\right|^2 \quad (6)$$

Equation (6) can be expanded into the envelopes and phases of the spatio-temporal modes:

$$P_{C,D}^{1,1}(\tau_0,\tau) = \frac{1}{4}\varepsilon_A^2(\tau_0+\tau)\varepsilon_B^2(\tau_0) + \frac{1}{4}\varepsilon_A^2(\tau_0)\varepsilon_B^2(\tau_0+\tau) - \frac{1}{2}\varepsilon_A(\tau_0)\varepsilon_B(\tau_0)\varepsilon_A(\tau_0+\tau)\varepsilon_B(\tau_0+\tau)\cos[\varphi_A(\tau_0) - \varphi_B(\tau_0) - \varphi_A(\tau_0+\tau) + \varphi_B(\tau_0+\tau)] \quad (7)$$

When the two photons come from the same input mode of the BS, the input states are given by $|\psi_{in}\rangle_{A,B} = a_A^\dagger a_A^\dagger|0,0\rangle_{A,B}$ and $|\psi_{in}\rangle_{A,B} = a_B^\dagger a_B^\dagger|0,0\rangle_{A,B}$. In this case, a similar evaluation is performed for $P_{C,D}^{2,0}(\tau_0,\tau)$ and $P_{C,D}^{0,2}(\tau_0,\tau)$.

### D. Model of the quantum beat between WCS

We consider that the WCS sources emit parallel-polarized photons with gaussian-shaped wave-packets in two well-defined frequency modes $\omega_A$ and $\omega_B$, described by

$$\xi_A(t) = \frac{1}{\sqrt[4]{\pi\sigma^2}} e^{-\left(t-\frac{\delta\tau}{2}\right)^2/(2\sigma^2)} e^{-j\left(\omega-\frac{\Delta}{2}\right)t}$$
$$\xi_B(t) = \frac{1}{\sqrt[4]{\pi\sigma^2}} e^{-\left(t+\frac{\delta\tau}{2}\right)^2/(2\sigma^2)} e^{-j\left(\omega+\frac{\Delta}{2}\right)t} \quad (8)$$

where $\omega = (\omega_A + \omega_B)/2$, $\sigma$ is the half-width at 1/e of the wave-packet, and $\delta\tau$ is the relative delay between the photons at the BS input. The frequency difference between the WCS sources, $\Delta = \omega_B - \omega_A$, is fixed. The squared-envelope absolute value integrates to unity from $-\infty$ to $\infty$.

The coincident detection probability of eq. (7) is solved using eq. (8), resulting in

$$P_{M,N}^{1,1}(\tau_0, \tau, \delta\tau) = \frac{1}{2\pi\sigma^2} e^{-\frac{\delta\tau^2/2+\tau^2}{\sigma^2}} e^{-\frac{2\tau_0^2+2t_0\tau}{\sigma^2}} \left[\cosh\left(\frac{\tau\delta\tau}{\sigma^2}\right) - \cos(\tau\Delta)\right] \quad (9)$$

The equation is then integrated over all values of $\tau_0$, resulting in the joint detection of photons with time difference $\tau$ at the output modes of the BS:

$$P_{M,N}^{1,1}(\tau, \delta\tau) = \frac{\sqrt{\pi}}{4\sqrt{2}\pi\sigma} e^{-\frac{\tau^2}{2\sigma^2}} e^{-\frac{\delta\tau^2}{2\sigma^2}} \left[\cosh\left(\frac{\tau\delta\tau}{\sigma^2}\right) - \cos(\tau\Delta)\right] \quad (10)$$

We also integrate over all values of $\delta\tau$ to account for the continuous-wave nature of the WCS sources, resulting in

$$P_{M,N}^{1,1}(\tau) = \frac{1}{2}\left[1 - e^{-\tau^2/(2\sigma^2)}\cos(\tau\Delta)\right] \quad (11)$$

Equation (11) exhibits an interference behavior depending on the coherence length of the states and on the frequency mismatch between the input photons at both ports.

When both photons reach the BS at the same port, second equation, they are randomly distributed to the output modes, with fixed probability ½ (in our CW case), i.e.,

$$P_{M,N}^{2,0}(\tau) = P_{M,N}^{0,2}(\tau) = \frac{1}{2} \quad (12)$$

The overall coincidence probability between the output modes 3 and 4 of the BS is given by summing the three elements in eqs. (11) and (12), weighted by (2): $P(1,1|\mu) = \mu^2 e^{-2\mu}$ and $P(2,0|\mu) = P(0,2|\mu) = \mu^2 e^{-2\mu}/2$. This accounts for the possibility of multi-photon emission by one source and vacuum by the other, bounded to a total of 2 photons. This results in the final expression for the coincidence probability which, after normalization by $P(1,1|\mu) + P(2,0|\mu) + P(0,2|\mu)$, results in

$$P_{coinc}(\tau) = \frac{1}{2} - \frac{1}{4} e^{-\tau^2/(2\sigma^2)}\cos(\tau\Delta) \quad (13)$$

### 4. Experimental setup

The experimental setup is composed by two main blocks, as shown in Fig. 2: preparation of two frequency-displaced WCSs with identical (faint) optical power and matched SOPs; and the acquisition of the interference pattern between these states in the HOM interferometer.

Here, we implement the WCSs from two uncorrelated versions of a CW signal split from an external cavity laser diode (LD). The self-heterodyne technique uses frequency and amplitude modulation to vary the difference between the optical frequencies of the WCSs by a controllable amount.

The optical signal passes through a variable optical attenuator (VOA$_1$) and is split in two arms by a symmetric beamsplitter (BS$_1$). The output modes of the BS$_1$ are decorrelated by a 8.5-km long optical fiber spool (OD$_1$) – a delay 80 times greater than the coherence length of the LD. Both arms are power balanced with VOA$_2$ and their SOPs are matched with polarization controller PC$_2$.

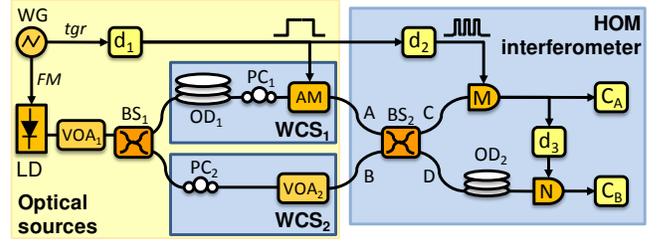

Fig. 2. Experimental setup for the proposed method. Frequency-displaced WCSs are created with a self-homodyne FM-based setup. LD: laser diode; WG: waveform generator; VOA: variable optical attenuator; d: delay generator; OD: optical delay; PC: polarization controller; AM: amplitude modulator.

The LD is frequency modulated (FM) with a (symmetric) triangular waveform with modulation depth $A$ and period $T$ (322.6 μs). The optical path OD$_1$ delays the output of WCS$_1$ and WCS$_2$ by an amount of time$\tau$, so that during part of the time the optical frequencies of both sources are swept linearly with a constant difference $\Delta = 2A\tau/T$. The output trigger signal of the waveform generator (WG) is delayed and formatted by a delay generator (d$_1$) and sent to a LiNiO$_3$-based amplitude modulator (AM). The pulses open 30 μs temporal gates that select the output of WCS$_1$ letting pass only photons whose frequency has a constant offset $\Delta$ to WCS$_2$ ones. This means that only the selected spectral range is allowed at the AM arm. The frequency difference ($\Delta$) between photons emerging from the two arms can thus be controlled by a proper choice of $A$ and $T$. In our case, we kept $T$ fixed for triggering reasons and varied the modulation depth $A$.

Photons from WCS$_1$ and WCS$_2$ are then recombined in a second symmetrical beamsplitter (BS$_2$). The beat spectrum between the emulated frequency-displaced optical sources is verified at bright power levels with an electrical spectrum analyzer (ESA) placed at one output mode of BS$_2$ (not depicted in Fig. 2).

The HOM interferometer employs two InGaAs APD-based SPDs operating in gated Geiger mode, one at each output mode of BS$_2$. The detectors have 15% detection efficiency and the width of their detection gate windows is set to 2.5 ns. SPD M is gated by a train of pulses at 1 MHz (d$_2$) within the 30-μs wide enable pulse (also sent to AM). A 100-m long optical delay line (OD$_2$) is placed before SPD N to allow for a gate delay scan around the matched temporal mode, performed with the delay generator d$_3$. Pulse counters (C$_M$ and C$_N$) acquire the photon-counting statistics of the heralded signal.

### 5. Results

Figure 3 shows the interference pattern measured for different frequency mismatches between the WCSs, ranging from zero to 200 MHz, with 40 MHz steps.

The figure shows the quantum beat frequency with the gaussian envelope of the mutual coherence time. The interference patterns were normalized to the coincidence count values measured with mismatched temporal modes. This condition of fully-distinguishable photons occurs at delay values greater than the mutual coherence time of the WCSs, outside the HOM dip.. Data was fit with the model presented in eq. (10) and the parameters $\sigma$ and $\varDelta$ were extracted.

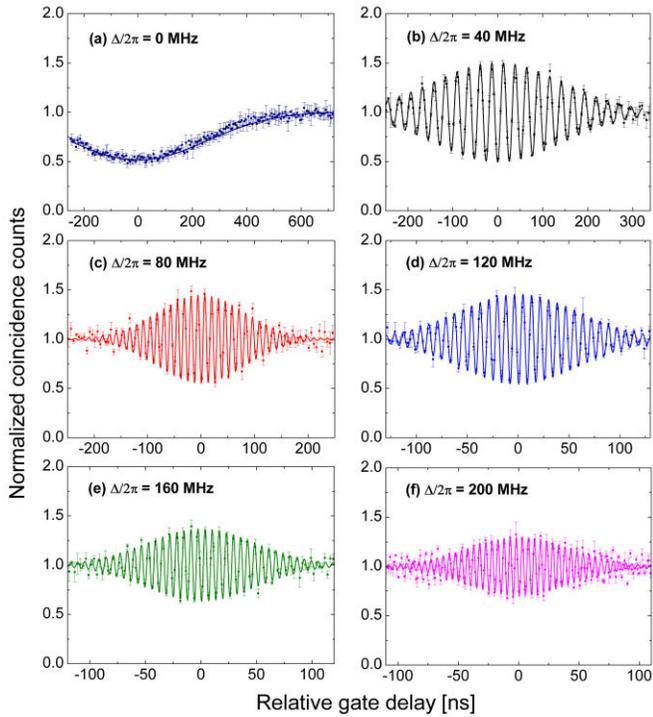

Fig. 3. Interference pattern of the WCS frequency-displaced by (a) 0 MHz, (b) 40 MHz, (c) 80 MHz, (d) 120 MHz,(e) 160 MHz, (f) 200 MHz. Error bars represent the statistical fluctuation of the measurement and the lines correspond to the theoretical fit. The envelope width gets narrower from (a) to (f) due to the enlarged linewidth of the laser sources.

The classical beat spectra were acquired for each configuration of the WCSs frequency mismatch, and are shown in Fig. 4. The classical beat notes measured with the ESA were fit with a gaussian model. The gaussian function was chosen here to match the prior description of the line-shape of the wave packet in our model. Distortion in experimental data appears due to imperfections during the emulation of the spectral lines. The central frequency and the linewidth parameters were extracted and compared to the values obtained from Fig. 3.

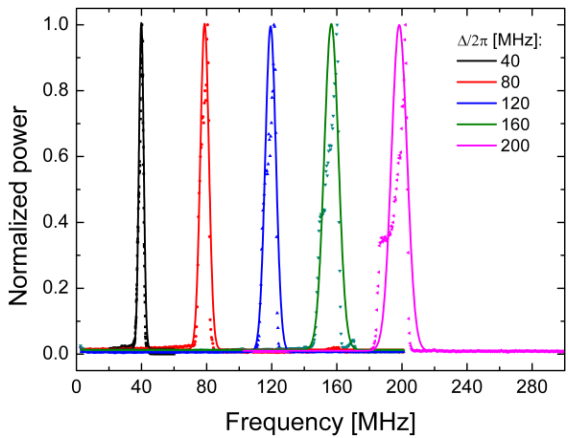

Fig. 4. Beat spectrum acquired with the ESA for bright versions of the frequency-mismatched optical sources. Lines correspond to gaussian fit.

The comparison results are depicted in Fig. 5, where we assess the equivalence of both techniques.

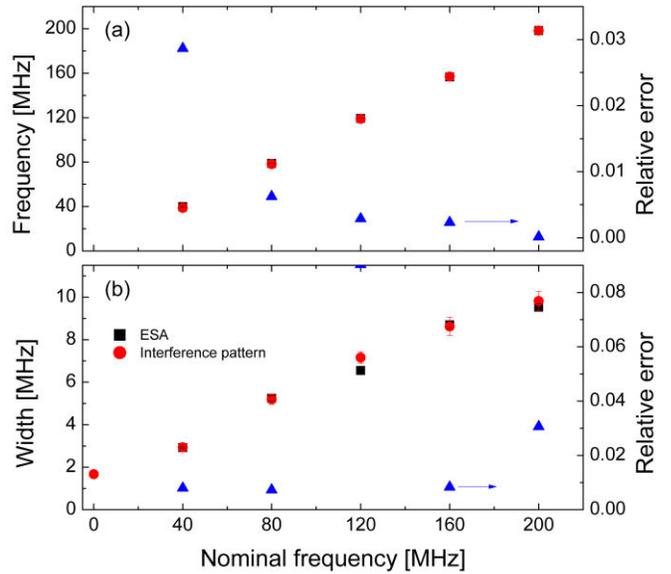

Fig. 5. (a) Frequency mismatch between the WCSs and (b) line width values (half-width at 1/e²) obtained by fitting the model to the interference pattern (red dots) and by fitting a gaussian to the ESA spectra (black squares). The relative error between the values is shown for each case (blue triangles). Error bars represent the uncertainty in the fit parameters.

The results for the frequency displacement agree within a relative error (computed as the absolute difference divided by the average between both values) smaller than 3%, getting better than 0.5% for the higher settings. Although the set of the linewidth value has not been controlled for each frequency condition, the results between both techniques agree with relative error within 3% for all range, except for the 120 MHz point, which seems to be an outlier. The linewidth of frequency-displaced optical source gets wider due to the increased slope of the triangular wave used in FM.

The central value of the frequency mismatch and the spectral width are displayed in the correlation plot of Fig. 6. The angular and linear coefficients of the linear fit are 1.00339 and 0, respectively.

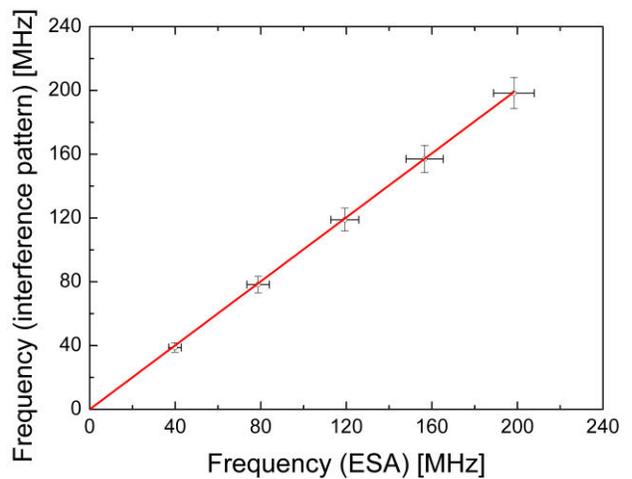

Fig. 6. Correlation between the beat frequency obtained from the interference pattern and with the ESA. Error bars represent the linewidth values (half-width at 1/e²). Red line is the linear fit of the data.

The characterization technique by coincidence counting depends on hardware features, as the resolution and step-size of the delay generator used in the HOM interferometer. A more fundamental limiting factor is related to the mutual coherence between sources and their spectral separation. Depending on the (lack of) coherence of the WCSs, the visibility of the interference pattern can fade for higher frequency mismatch. This issue can be circumvented provided a tunable laser source is used, so the probing laser line can be positioned spectrally close to the test WCS.

Another fundamental limitation concerns the time-resolution of the measurement [8,14]. The temporal width of the detection gate of the SPDs must be smaller than the oscillation of the beat note to be measured, otherwise the interference pattern is averaged out and the information related to the frequency mismatch could be lost.

### 6. Conclusions

We have demonstrated a method for the spectral characterization of coherent states in the weak regime based on two-photon interference in a beamsplitter. A WCS source, under test, is fed into a beamsplitter together with a reference WCS, and the interference pattern is obtained through coincidence counts in a *Hong-Ou-Mandel* interferometer. The parameters are extracted through the fit of the theoretical model for the two-photon interference, revealing the frequency mismatch and the convolved coherence length of the sources. The method was validated when compared to the spectrum obtained from the optical beat of bright versions of the optical sources in a photodiode, observed in an electrical spectrum analyzer. The results show the equivalence between both techniques for different frequency mismatch values between optical sources.